\documentclass[useAMS,usenatbib]{mn2e}

\usepackage{graphicx}
\usepackage{setspace}
\usepackage{natbib}
\usepackage{color}
\usepackage{amsmath,amssymb}
\usepackage{times}
\usepackage{aas_macros}
\usepackage{multirow}

\title[Stellar streams around the Magellanic Clouds]{Stellar streams
  around the Magellanic Clouds}

\author[Belokurov and Koposov]{Vasily Belokurov$^{1}$\thanks{E-mail:vasily@ast.cam.ac.uk} and Sergey E. Koposov$^{1}\thanks{E-mail:koposov@ast.cam.ac.uk}$
\\ $^{1}$Institute of Astronomy, Madingley Rd, Cambridge, CB3 0HA}

\begin{document}


\maketitle

\label{firstpage}

\begin{abstract}
Using Blue Horizontal Branch stars identified in the Dark Energy
Survey Year 1 data, we report the detection of an extended and lumpy
stellar debris distribution around the Magellanic Clouds. At the
heliocentric distance of the Clouds, overdensities of BHBs are seen to
reach at least to $\sim30^{\circ}$, and perhaps as far as
$\sim50^{\circ}$ from the LMC. In 3D, the stellar halo is traceable to
between 25 and 50 kpc from the LMC. We catalogue the most significant
of the stellar sub-structures revealed, and announce the discovery of
a number of narrow streams and diffuse debris clouds. Two narrow
streams appear approximately aligned with the Magellanic Clouds'
proper motion. Moreover, one of these overlaps with the gaseous
Magellanic Stream on the sky. Curiously, two diffuse BHB
agglomerations seem coincident with several of the recently discovered
DES satellites. Given the enormous size and the conspicuous lumpiness
of the LMC's stellar halo, we speculate that the dwarf could easily
have been more massive than previously had been assumed.
\end{abstract}

\begin{keywords}
Galaxy: fundamental parameters --- Galaxy: halo --- Galaxy: kinematics
and dynamics --- stars: blue stragglers --- stars: horizontal branch
\end{keywords}

\section{Introduction}

\begin{figure*}
  \centering
  \includegraphics[width=\textwidth]{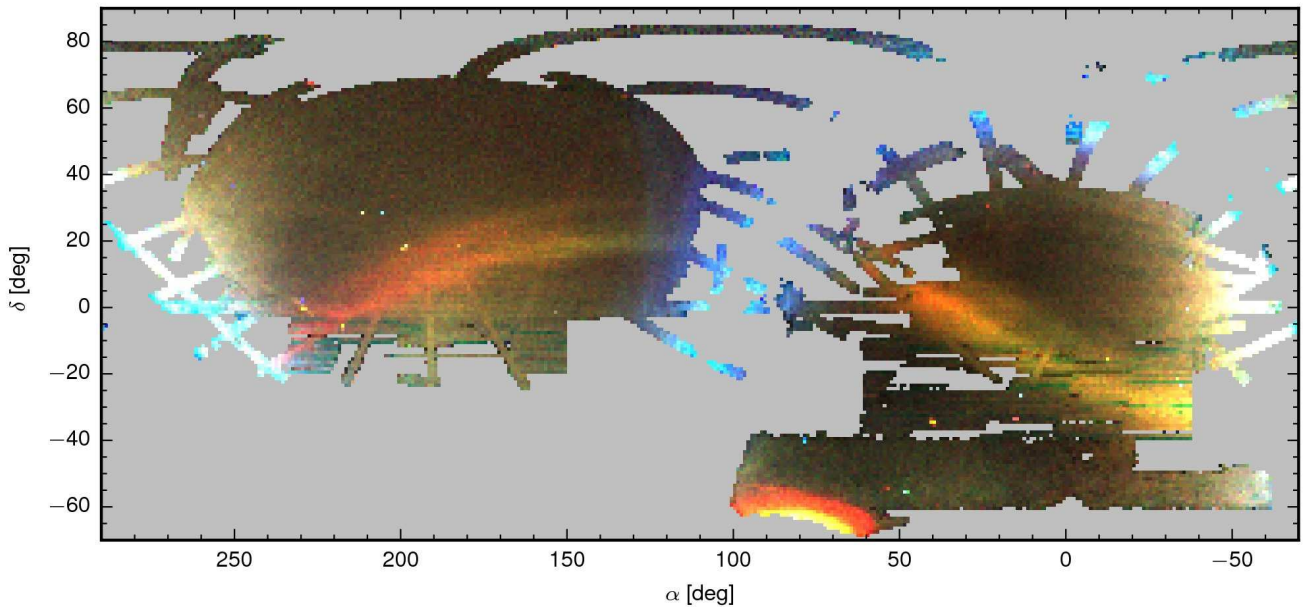}
  \caption[]{\small Jigsaw puzzle of the Galactic stellar halo as
    traced by the MSTO stars. This false-colour composite image
    combines data from three different surveys: SDSS DR9, VST ATLAS
    and DES Year 1. R, G and B channels of the image are greyscale
    density maps of stars with $0.2<(g-i)<0.5$ within the following
    magnitude ranges: $22>i>21$ (red), $21>i>20$ (green) and $20>i>19$
    (blue). The Sgr stellar stream and the LMC (with the centre at
    $\alpha=80\fdg9$ and $\delta=-69\fdg7$) are the most
    prominent halo structures visible. Note also two faint and fuzzy
    overdensities of approximately green colour at RA$\sim100^{\circ}$
    and RA$\sim30^{\circ}$. These must lie much closer along the line
    of sight as compared to the LMC.}
  \label{fig:stitch}
\end{figure*}

\begin{figure*}
  \centering
  \includegraphics[width=0.98\textwidth]{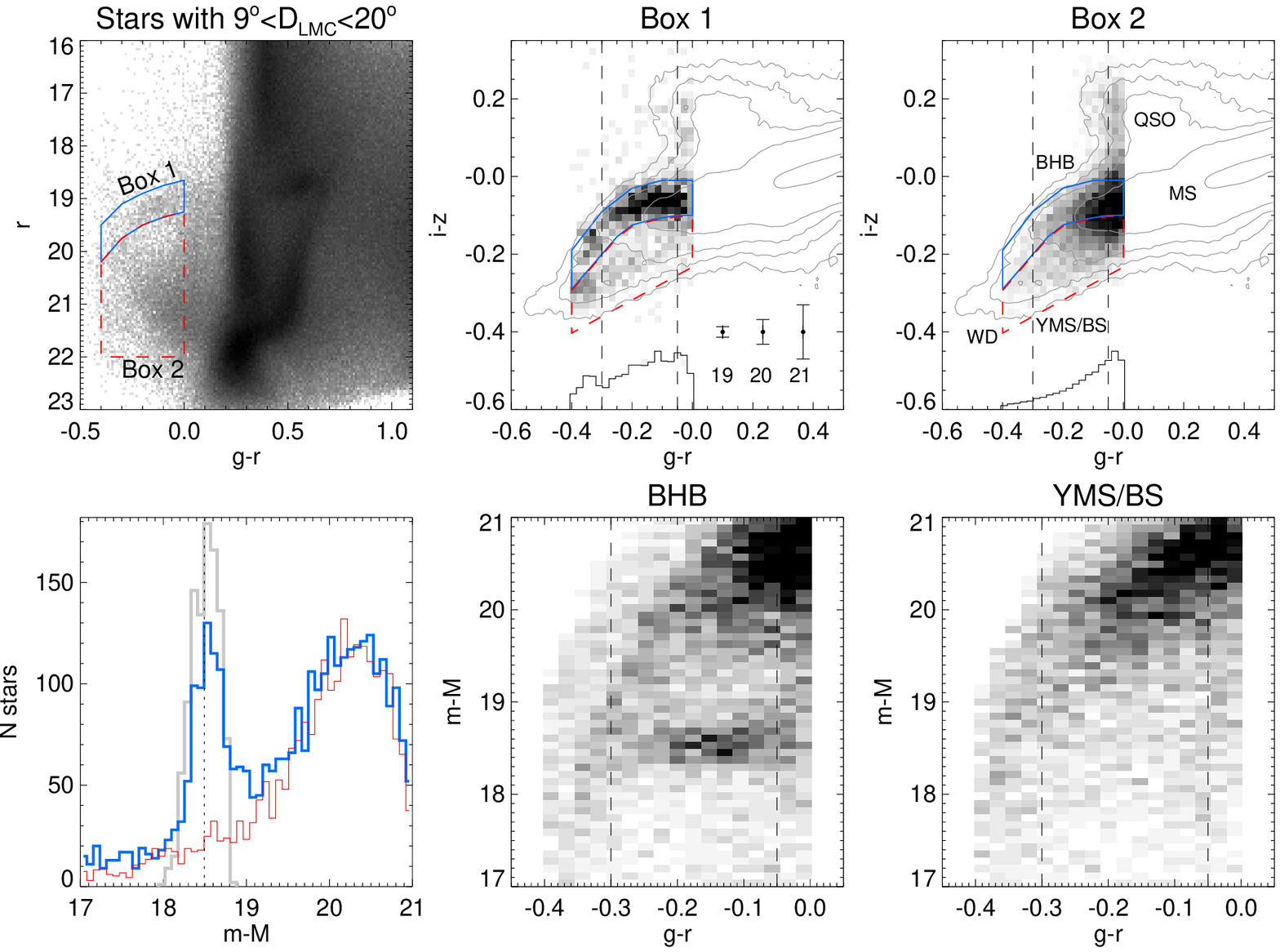}
  \caption[]{\small Blue Horizontal Branch star identification with
    the DES's $griz$.  \textit{Top left:} Hess diagram (CMD density)
    of all stars with $9^{\circ} < D_{\rm LMC} < 20^{\circ}$. Familiar
    features such as the Main Sequence Turn-Off, the Red Giant Branch
    and the Red Clump are visible in the red portion of the CMD
    ($g-r>0$). In the blue, two features stand out: the Blue
    Horizontal Branch (at $r < 20$) and the cloud of Young Main
    Sequence and Blue Straggler stars (with $r>20$). Blue (red) lines
    mark the selection Box 1 (Box 2) used to isolate the BHB (YMS/BS)
    stars. \textit{Top centre:} Density in the $g-r$, $i-z$ plane for
    stars selected to lie in the Box 1 in the top left panel. Contours
    give the density distribution of all stars. A tight sequence of
    genuine BHB stars is visible, curving from $g-r\sim 0$, $i-z\sim
    -0.1$ to $g-r\sim -0.4$, $i-z\sim-0.3$. Blue (red) box shows the
    colour-colour decision boundary used to select BHB (YMS/BS)
    stars. The coordinates of the vertices of the selection boxes are
    given in Tables~\ref{tab:cmd_bhb} and ~\ref{tab:cmd_bs}. Vertical
    dashed lines show the restricted range of $g-r$ used for the BHB
    selection reported here. The histogram gives the $g-r$
    distribution of the BHB stars that passed both the CMD and
    colour-colour cuts. Three points with error-bar demonstrate the
    behaviour of $i-z$ error as a function of $r$ for stars with
    $g-r<0$ and $i-z < 0$. \textit{Top right:} Same as the top centre
    panel, but for the stars in Box 2. Portions of the $g-r, i-z$
    space are marked with the names of objects dominating the density
    at that location. \textit{Bottom left:} Distance modulus
    distribution. The blue (red) line gives the histogram of $m-M$
    calculated using a version of equation 7 of \citet{De11}, for all
    stars with $9^{\circ} < D_{\rm LMC} < 20^{\circ}$ lying in the
    blue (red) box shown in the top centre and top right panels. The
    blue curve corresponds to the LMC's BHB stars and peaks at
    $m-M\sim 18.5$ in perfect agreement with the recent measurement by
    \citet{LMCdist}. Grey line is the histogram of BHB candidate stars
    selected using both CMD cuts shown in top left panel as well as
    the colour-colour cuts shown in the center and right
    panels. \textit{Bottom center:} BHB distance modulus as a function
    of the $g-r$ colour. Density of stars selected using the blue box
    from the panel directly above is shown. Notice the growing
    contamination from the YMS/BS stars outside the region marked with
    dashed lines. \textit{Bottom right:} Same as previous panel, but
    for the YMS/BS stars selected using the colour-colour box shown in
    red.}
  \label{fig:bhb}
\end{figure*}

This Universe is conjectured to have assembled hierarchically, from
the bottom up, yet the evidence for accretion onto dwarf galaxies with
luminosities $\leq 10^{10} L_{\odot}$ is currently scarce
\citep[][]{rich2012, md2012, amorisco2014}. In the cosmic pecking
order, the Magellanic Clouds are just a position down from the Milky
Way and, hence could, in principle, given their proximity, serve as a
typical example of the sub-L$_*$ satellite galaxy assembly in the
$\Lambda$CDM Cosmology. The extent and the amount of lumpiness in the
stellar halo (if it exists) of the Large Magellanic Cloud (LMC) would
not only present another crucial piece of evidence that our structure
formation theory works on all scales; it would also give us a close-up
view of the emergence and the destruction of sub-structure within
sub-structure. Moreover, the LMC's dark matter (DM) distribution as
well as the galaxy's orbital history could be accurately
constrained. However, to date, the details of the gravitational
interaction between the LMC and the SMC remain muddled, and precious
little information is available as to the signs of accretion of even
smaller fragments onto either of the Clouds.

The gravitational interaction between the two Clouds has been under a
close scrutiny ever since the Magellanic gaseous Stream (MS) was
discovered \citep[][]{wannier1972,matheson1974}. First attempts at
modelling the MS inaugurated a new promising sub-field of Galactic
Astronomy, the Stream Dynamics, yielding new interesting constraints
on the DM halo of the Milky Way
\citep[e.g.][]{davies1977,lin1977,lin1995} as well as its gaseous halo
\citep[][]{meurer1985,moore1994,heller1994,mas2005} depending on the
mechanism assumed responsible for the production of the Stream. Very
quickly it was established that the Stream ought to originate from the
SMC rather then the LMC \citep[][]{murai1980,connors2006}, thus
bringing into light the complex history of three-body encounters
between the Clouds and the Galaxy. Even though the time the Clouds
have spent together is not well constrained, several MS simulations
revealed that it was possible for the LMC and the SMC to become a pair
only recently as most of the salient MS features could be reproduced
with a 2 Gyr orbital history \citep[e.g.][]{gardiner1996}. As the
observational data on the gas in the MS and the Clouds themselves grew
in volume and complexity
\citep[][]{putman2003,hz2006,nitya2006a,nitya2006b,nidever2008,nidever2010,nitya2013},
so did the numerical models
\citep[][]{besla2007,besla2010,diaz2011,diaz2012}.

As every tidal model of the MS production predicts a stellar
counterpart to the observed gaseous component, several searches were
conducted to uncover stars coincident with the Stream
\citep[][]{sanduleak1980,rc1982,brueck1983,mcmahon1989,kunkel1997,guhathakurta1998}. Intriguingly,
none of them yielded a convincing detection. The null detection could
either be viewed as an argument against the tidal nature of the MS, or
explained by the extremely low surface brightness levels of the
stellar debris, actually in agreement with most simulations
\citep[][]{weinberg2000,diaz2012,besla2013}. Additionally, as
\citet{diaz2012} argue, the gaseous stream produced by stripping of
the disk and the stellar stream originating (at least in part) from
the SMC's old spheroid could be misaligned.

\begin{figure}
  \centering
  \includegraphics[width=0.45\textwidth]{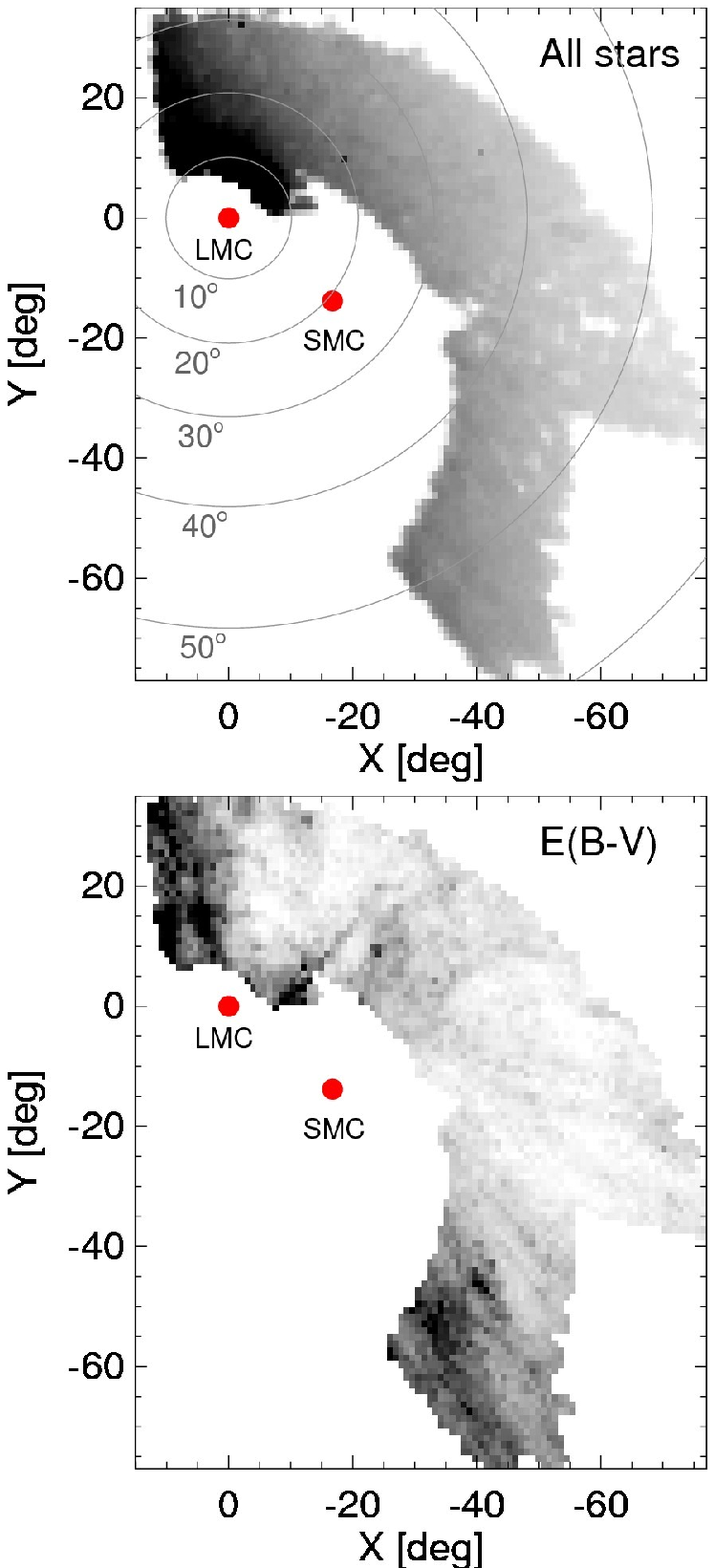}
  \caption[]{\small \textit{Top:} Density of all stars in the DES
    Year 1 dataset in gnomonic projection with the tangent point at
    the LMC's location. Filled red circles mark the locations of the
    LMC and the SMC. Lines of equal spherical distance from the LMC
    are also shown. This map is 100 pixels on the
    side. \textit{Bottom:} Distribution of the Galactic reddening in
    the survey's footprint. The map has the same dimensions as that
    shown in the left panel. The greyscale is set to vary between the
    2nd and the 98th percentiles of the $E(B-V)$ values in the area,
    i.e. 0.0075 and 0.075.}
   \label{fig:dendust}
\end{figure}

Seemingly more successful have been the attempts to map out the
stellar halo of the LMC. Bits of the evidence in favour of an extended
spheroidal component come from the kinematics of planetary nebulae and
RR Lyrae \citep[e.g.][]{feast1968,minniti2003}. On the other hand,
globular clusters have been shown not to be part of a halo
\citep[][]{freeman1983}. Star count maps built with a variety of
tracers have helped to detect shreds of an extended structure beyond
$10^{\circ}$ (and as far as 15 kpc) from the LMC
\citep[e.g.][]{irwin1991,kinman1991}. However, according to
\citet{alves2004}, these measurements are also consistent with an
extended disk model. Several studies have gone as far as
$\sim20^{\circ}$ from the Cloud. For example,
\citet{majewski1999,majewski2009} report a kinematic detection of the
LMC red giant branch (RGB) stars at around 23$^{\circ}$. Similarly,
\citet{munoz2006} traces RGB stars consistent with the origin in the
LMC to the location of the Carina dwarf, i.e. $22^{\circ}$ from the
LMC. This detection is confirmed by \citet{mcmonigal2014} with the
help of a matched filter technique applied to the deep follow-up
imaging around Carina. Unfortunately, yet again, as shown most
recently by \citet{Mackey2015}, rather than a genuine halo population,
these stellar debris could be a part of the large stream-like feature
that appears to be torn off the LMC's disk.

Perhaps the most comprehensive approach to date to surveying the
periphery of the Magellanic System is exemplified by the Outer Limits
Survey \citep[][]{saha2010}, which gathered deep multi-band photometry
over tens of square degrees around the Clouds. The OLS reports
unambiguous identification of LMC's Main Sequence stars out to
16$^{\circ}$ (from the LMC). Moreover, the star counts appear to
follow an exponential profile. As the authors point out, the most
straightforward interpretation of this result is that the LMC's disk
stretches as far as 10 scale lengths. However, according to
\citet{saha2010}, if the mixture of the LMC's structural components
looked anything like that of the Milky Way, the disk would anyway
dominate the star counts to at least 10 scale lengths, or
further. Therefore, even though the OLS does not detect the LMC's
stellar halo, it can not confirm its absence either.

As illustrated above, the quest to track down stellar debris at large
distances from the LMC (and the SMC) appears to have been frustrated
by the lack of deep, wide and {\it continuous} coverage of the
outskirts of the Magellanic Clouds. Motivated by this, we choose to
scout for the Magellanic stellar halo sub-structure using Blue
Horizontal Branch stars selected from the Dark Energy Survey (DES)
Year 1 public dataset. First, the DES delivers an uninterrupted view
of the relevant patch of the Southern sky that is an order of
magnitude larger in area compared to all previous surveys, while
reaching depths similar to the most ambitious of the earlier
attempts. The choice of BHBs as tracers is rather obvious too: these
old and metal-poor stars suffer little contamination from other
stellar populations, can be easily picked up as far as 100 kpc or
beyond, and, finally, are one of the best stellar standard rulers
available. This paper is organised as follows. To begin with, we
present a panoramic view of the Milky Way stellar halo in
Section~\ref{sec:bigger}. This is followed by presentation of the
details of the BHB selection process in Section~\ref{sec:data}. We
describe the sub-structure uncovered in
Section~\ref{sec:streams}. Finally, Section~\ref{sec:disc} summarizes
our discoveries and puts them into context.


\section{The Bigger Picture. The Galaxy's stellar halo}
\label{sec:bigger}

The Magellanic Clouds are on their way to merge with the already
crowded Milky Way halo. To place the subsequent analysis into
perspective, we create a mosaic map of the Galactic stellar halo in
equatorial coordinates, shown in Figure~\ref{fig:stitch}. This
panoramic view of the Galaxy's outskirts is put together using MSTO
stars from three different surveys: SDSS DR9 \citep[][]{sdssdr9}, VST
ATLAS \citep[][]{vstatlas} and DES Year 1 \citep{Koposov2015}. This
false-colour composite image shows density distributions of stars with
$0.2<(g-i)<0.5$ in three different magnitude ranges. The red channel
corresponds to the density of MSTO stars with $21<i<22$ (most distant
stars), the green channel gives the density of stars with $20<i<21$,
and the blue channel corresponds to the nearest MSTO stars with
$19<i<21$.

The two most prominent structures in this map are the Sagittarius
stream \citep[see e.g.][]{Ma03, Be06, Ko12, Belokurov2014}, looping
through the entire sky covered by the SDSS and the VST ATLAS, and the
LMC at $100^{\circ} >$ RA $> 50^{\circ}$ at the bottom of the DES Year
1 footprint. Even though many other familiar halo sub-structures are
discernible elsewhere in the image, we limit our attention to several
new overdensities visible in the area of the sky covered by the DES
survey. Seemingly connected to the LMC disk at RA$\sim100^{\circ}$, a
faint and almost vertical tail of MSTO stars is noticeable. It is
possible though that rather than being a genuine halo substructure,
this overdensity is related to the disk instead, since its galactic
latitude is $b\sim-25^{\circ}$. Additionally, there is a fuzzy and
faint green overdensity of stars at around RA$\sim30^{\circ}$. This
stellar cloud has been recently reported as Eridanus-Phoenix
overdensity by \citet{li2015}.

Given their colour in Figure~\ref{fig:stitch}, the two DES MSTO
overdensities mentioned above are located much closer along the line
of sight as compared to the LMC. Their genealogy is yet unknown, but
their proximity to the Magellanic Clouds might turn out to be purely
coincidental. Note that apart from these two rather faint
overdensities and the LMC itself, the area covered by the DES Year 1
data appears rather quiet and unremarkable, as traced by the MSTO
stars. However, the BHBs tell a very different story.

\section{Blue Horizontal Branch star identification with DES}
\label{sec:data}

To study the stellar halo sub-structure around the Magellanic Clouds,
we use the photometric catalogs obtained from the publicly released
DES Year 1 imaging \citep[see][for details]{Koposov2015}, in
particular, the latest improved version of the reduction as reported
in \citet{Mackey2015}. While the calibration of the $gri$-band data
has been described in \citet{Koposov2015}, the calibration of the
$z$-band data requires a separate discussion, as it was done
differently to the $gri$ photometry. The reason is the lack of
$z$-band measurements in the APASS (the AAVSO Photometric All-Sky
Survey) catalogue. Therefore we compute the APASS $z$-band magnitude
using the $g$ and $r$ APASS magnitudes. More precisely, we describe
the SDSS (and therefore also APASS) $z$-band magnitude as a 4-th
degree polynomial of the $g-r$ colour \footnote{$z=-0.2126+1.2227(g-r)
  -4.1545(g-r)^2+ 4.0484(g-r)^3 -1.4355(g-r)^4$}. The limiting
magnitude in the $z$ band is $\sim 21.9$.

DES provides the deepest and widest panorama of the environs of the
Magellanic Clouds to date. The survey's Year 1 footprint covers only a
portion of the halo of the LMC and an even smaller segment of the
SMC's outskirts. However, the DES's un-interrupted view is crucial
when making sense of such a large, diffuse and fragmented
structure. Our catalogs cover $\sim2200^{\circ}$ square degrees just
North of the Clouds and contain $griz$ magnitude measurements
(tied to the SDSS/APASS photometric system) for $\sim1.5 \times
10^8$ stars with $g < 23.5$. In the analysis that follows, the
photometry is corrected for the effects of extinction using the dust
maps of \citet{SFD} and the conversion suggested by
\citet{Schlafly2011}. For the star/galaxy separation, we use the
following conservative cut: $|$SPREAD\_MODEL$| < 0.002$ simultaneously
in three bands, $g$, $r$ and $i$. Note that at faint magnitudes, this
morphological classification is less complete, but more pure as
compared to the criterion used in \citet{Koposov2015}.

\begin{figure*}
  \centering
  \includegraphics[width=0.98\textwidth]{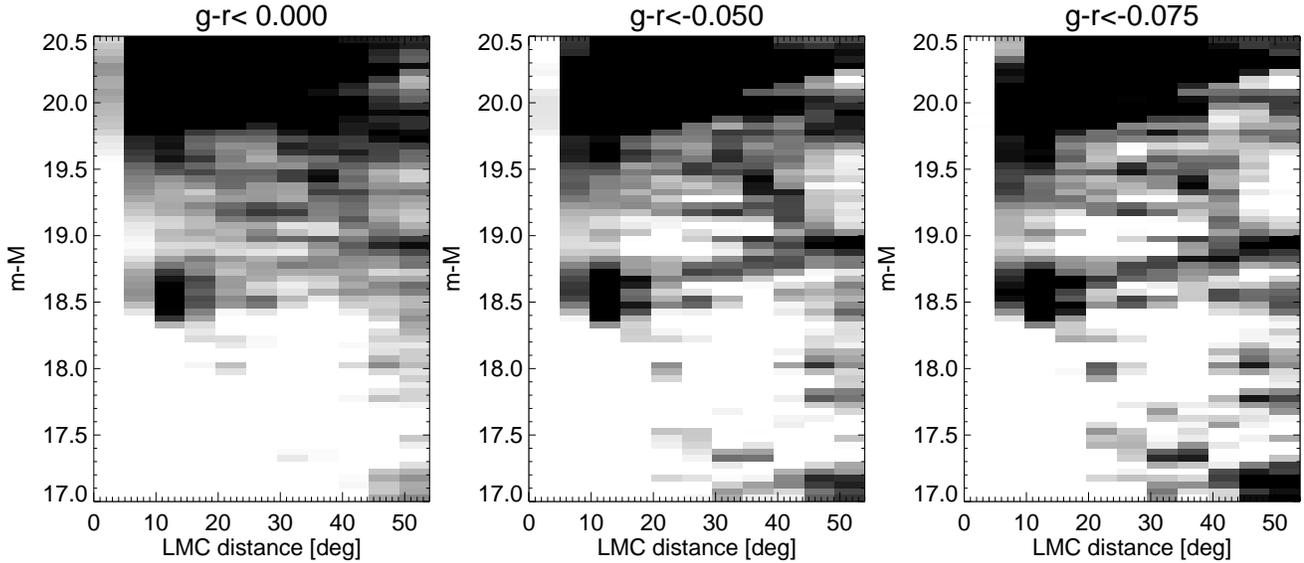}
  \caption[]{\small Stellar sub-structure around the Magellanic Clouds
    as traced by BHBs. Each panel shows the density of the BHB
    candidate stars in the plane of distance modulus and the angular
    distance from the LMC. The maps are 11$\times$70 pixels, and are
    smoothed with a Gaussian kernel with a FWHM of 1.7 pixels. From
    left to right, the red $g-r$ cut used for the BHB selection (see
    Figure~\ref{fig:bhb}) is lowered from 0 to -0.075, as indicated at
    the top of each panel. Limiting the BHB candidate selection to
    bluer $g-r$ colours helps to improve the effects of the YMS/BS
    contamination. The contamination is most severe in the
    distribution shown in the left panel. As the YMS/BS contamination
    becomes reduced (middle panel), several dark (corresponding to an
    enhanced BHB density) sequences are clearly visible starting at
    $m-M\sim 18.5$, i.e. the LMC's heliocentric distance. The signal
    strengths improves as the BHB sample becomes purer (right panel):
    the most prominent dark band corresponds to the BHB debris whose
    heliocentric distance increases with angular distance from the
    LMC. As obvious from the last two panels, the LMC's stellar halo,
    as traced by BHBs, extends out to $\sim 50^{\circ}$ from the
    dwarf. Note that the DES coverage is very sparse near the LMC
    center, however, at distances larger than 10$^{\circ}$ more than a
    quarter of the LMC halo is sampled by DES.}
   \label{fig:grcuts}
\end{figure*}

BHBs are a go-to stellar tracer for the Galactic halo studies
\citep[see e.g.][]{Ya00, Ne03,Xue2008,Bell2010,Ru11,De11}. In the
halo, not only can these old and metal-poor stars be easily identified
above the foreground of other populations thanks to their peculiar
colour, they are also one of the best stellar distance estimators
available, outperformed only by their immediate neighbors on the
Hertzsprung-Russell diagram, the RR Lyrae. Thanks to their unique
properties, BHBs have proven to be a powerful tool to scrutinize the
Galactic halo out from the core to its far-flung fringes
\citep[e.g.][]{Belokurov2014,Deason2014}.

As the name suggests, the spectral energy distribution of a BHB star
peaks at short wavelengths. There, in the blue, minute changes in the
shapes of the Balmer lines accumulate to provide a powerful diagnostic
to distinguish BHBs from other stars with similar temperatures
\citep[see e.g.][]{Sirko2004}. Impressively, these signatures can also
be picked up by means of broad-band photometry only \citep[see
  e.g.][]{Ya00}. Provided an accurate measurement of near-UV flux
excess exists, e.g. SDSS's $u-g$ colour, BHBs immediately break away
from quasars and White Dwarfs (WD).  Differences between Young Main
Sequence stars (YMS) or similarly looking Blue Stragglers (BS) and
BHBs are more subtle and typically require higher quality $u$-band
photometry \citep[see e.g.][]{Deason2012}. DES, however, does not
observe in the $u$ band, hence the BHB identification techniques
described above are not applicable. Nevertheless, as shown by
\citet{Vickers2012}, near-IR photometry can be used instead with an
almost equally impressive success. According to the authors, rather
than the Balmer series, the Paschen lines serve as the discriminator,
resulting in the noticeable differences in the $i-z$ colour of BHBs,
QSO, WDs and YMS/BS stars.

Figure \ref{fig:bhb} gives the details of the BHB selection process as
applied to the DES data. Top left panel of the Figure shows the Hess
diagram (i.e. the density of stars in the colour-magnitude space) of
all stars with angular separations between 9$^{\circ}$ and
20$^{\circ}$ from the center of the LMC, which we assume to have
coordinates $\alpha=80\fdg8937$ and $\delta=-69\fdg7561$
\citep[][]{Alan2012}\footnote{Note, however, that the dynamical centre
  of the LMC has recently been estimated to lie at $\alpha=78\fdg76$
  and $\delta=-69\fdg19$ by \citet[][]{vdm2014}}. At these distances,
  the Colour-Magnitude Diagram (CMD) contains plenty of both young and
  old populations, most easily discernible at $g-r<0$. The selection
  Box 1 (Box 2) used to isolate the BHB (YMS) candidate stars are
  shown in blue (red) respectively. It is clear from this CMD, that
  even though the BHBs dominate at brighter apparent magnitudes, the
  YMS/BS contamination of the area shown in blue is not
  negligible. The greyscale density of the thus selected BHB
  candidates in the plane of $g-r$ and $i-z$ colours can be seen in
  the top middle panel of the Figure. The narrow and curving
  overdensity corresponding to the genuine BHBs is clearly
  visible. The boundaries of the BHB signature are demarcated by the
  box shown in blue. At bluer $i-z$ colours lie the YMS and BS stars,
  mixed with Galactic WDs - this region of $g-r$,$i-z$ plane is marked
  with a dashed red line. The coordinates of the $g-r$,$i-z$ selection
  boxes are given in Tables~\ref{tab:cmd_bhb} and ~\ref{tab:cmd_bs}.

\begin{table*}
\caption{Boundaries of the BHB selection box}
\begin{tabular}{|c c c c c c c c c c c c c c|}
\hline
$g-r$&-0.4 & -0.3 & -0.25 & -0.2 & -0.13 & -0.05 &  0.0 &  0.0 & -0.1 & -0.2 & -0.3 & -0.4 & -0.4 \\
$i-z$&  -0.3 &   -0.2 &   -0.2 &   -0.1 &   -0.1 &   -0.1 &   -0.1 &   -0.0 &   -0.0 &   -0.0 &   -0.1 &   -0.2 &   -0.3 \\
\hline
\end{tabular}
\label{tab:cmd_bhb}
\end{table*}
\begin{table*}
\caption{Boundaries of the YMS/BS selection box}
\begin{tabular}{|c c c c c c c c c c c|}
\hline
$g-r$&-0.4&-0.3&-0.25&-0.2&-0.13&-0.05&0.0&0.0&-0.4&-0.4\\
$i-z$&-0.3&-0.2&-0.2&-0.1&-0.1&-0.1&-0.1&-0.2&-0.4&-0.3\\
\hline
\end{tabular}
\label{tab:cmd_bs}
\end{table*}

The highly clustered behaviour of the BHBs in the colour-colour space
can be contrasted with the distribution of YMS/BS stars, as shown in
the top right panel of Figure~\ref{fig:bhb}. The MS and pseudo-MS
stars tend to congregate at redder $g-r$ colours and cover a much
broader range of $i-z$ colour. At the bottom of each of the two
panels, a histogram showing the corresponding $g-r$ distribution is
given to emphasise these differences. It appears beneficial to get rid
of the reddest (in $g-r$ sense) of the BHB candidates to minimise the
YMS/BS contamination. Further insight into the properties of the
selected BHB candidates can be gleaned from the panels in the bottom
row of the Figure. Here, distributions of the distance modulus values
of the BHB and YMS/BS samples are depicted, assuming the same BHB
absolute magnitude calibration for both. BHB's intrinsic luminosity is
primarily a function of its temperature, or equivalently, its $g-r$
colour. To this end, we use a slightly modified version of formula 7
of \citet{De11} to obtain estimates of $M_g$ and, hence,
$m-M$ \footnote{$M_g=0.398-0.392(g-r)+2.729(g-r)^2+29.1128(g-r)^3+113.569(g-r)^4$. This
  absolute magnitude is only $\sim$0.02 mag brighter than that of
  \citet{De11}, and hence all results presented here should remain
  unchanged if their calibration is used.}. Note, however, that as
pointed by \citet{Fe13}, the BHB's absolute magnitude also weakly
depends on its metallicity.

Blue (red) histogram in the bottom left panel of Figure~\ref{fig:bhb}
gives the distance modulus distribution for stars that fall into the
blue (red) box shown in the top middle and top right panels, i.e. the
BHB (YMS/BS) stars. Additionally, these are limited to lie in the same
range of angular distances as the stars displayed in the top left
panel, and have $-0.3 < g-r < -0.05$. At low $m-M$, the BHB
distribution peaks at $\sim18.5$ which is in perfect agreement with the
recent LMC distance estimate by \citet{LMCdist}. To illustrate the
completeness of the proposed BHB selection, underlying the blue
histogram, is the grey line showing the distribution of stars selected
as above but instead of the $griz$ cut, required to lie in the blue
box shown on the CMD in top left panel of the Figure. The difference
in height between the blue and grey peaks is predominantly due to the
polluting YMS/BS stars. The YMS/BS histogram, shown in red, shows no
appreciable amount of BHBs around the LMC distance modulus, and peaks
$\sim$2 mag further. Note that these large $m-M$ values do not
correspond to the true distances to these stars, but rather reflect
the fact that they are on average $\sim$2 mag fainter as compared to
the BHBs. The need for the additional $g-r$ cut as mentioned above
becomes clear on examination of the bottom middle panel. Here, the
behaviour of the $m-M$ as a function of the $g-r$ colour is shown for
the BHB candidates. Outside the $-0.3 < g-r < -0.05$ range, the BHB
and YMS/BS sequences start to merge and the contamination tends to
increase. Additionally, as obvious from the left panel of Figure 4 of
\citet{De11} and the middle panel of Figure 1 of
\citet{Belokurov2014}, outside this $g-r$ range, the BHB absolute
magnitude can deviate mildly from the ridgeline prescribed by the
polynomial fit.

Overall, as this Section demonstrates, for a particular $g-r$ range, a
simple $griz$ selection box can be used to obtain highly complete
samples of BHB stars. Such samples will always be contaminated with
YMS/BS stars, and this contamination grows for stars with redder $g-r$
colour. However, in the distance modulus space, the contaminants'
distribution always appears broader. Moreover, for co-distant
populations, this contamination peaks $\sim$2 mag fainter, and as
such, can be easily identified and discarded. Quite simply, there is
enough evidence to conclude that the asbolute majority of narrow peaks
in the $m-M$ should be attributed to the BHBs.

\section{BHB overdensities around the Magellanic Clouds}
\label{sec:streams}

As the DES Year 1 footprint provides a better view of the LMC's
surroundings as compared to the SMC's coverage, we choose to present
the maps of the BHB density distribution in a gnomonic projection
centred on the LMC \citep[similar to e.g.][]{Mackey2015}. The shape of
the survey's footprint in these coordinates can be seen in
Figure~\ref{fig:dendust}. The top panel of the Figure shows the
greyscale density distribution of all stars in the dataset. The
dramatic density enhancement in the left corner of the DES Year 1
field of view is partly due to the LMC's stars (at
$-10^{\circ}<X<10^{\circ}$ and $Y<15^{\circ}$) and partly due to the
Galactic disk (at $X\sim10^{\circ})$. While the footprint shape is
roughly rectangular in equatorial coordinates, it shows plenty of
curvature far away from the tangent point. In gnomonic projection, the
area is not strictly conserved, but suffers little distortion within
30$^{\circ}$ away from the LMC, as indicated by lines of equal angular
distance. Beyond that, there is a noticeable stretch, i.e. objects
appear further than they actually are. The footprint reaches only
slightly closer than 10$^{\circ}$ from the LMC centre. The bottom panel
of the Figure displays the distribution of the interstellar extinction
in the DES Year 1 footprint. The greyscale is set to vary between the
2nd and the 98th percentile of $E(B-V)$ values in the area, namely
0.0075 and 0.075. This corresponds to 0.025 and 0.25 magnitudes of
extinction in the $g$ band, and 0.0095 and 0.095 magnitudes in the $z$
band. The darker regions at either ends of the footprint show the
largest (relative) amount of dust reddening. This is because these
portions of the DES-covered area are closest to the Galactic disk.

Figure~\ref{fig:grcuts} illustrates how the Magellanic stellar halo
reveals itself as layers of YMS/BS stars are gradually removed. Here,
the density of the BHB candidate stars is shown in the plane of the
distance modulus and the angular distance from the LMC. From left to
right, the red $g-r$ cut, used for the BHB selection, is lowered from
0 to -0.075. As discussed above, the histograms in the top middle and
top right panels of Figure~\ref{fig:bhb} show that the BHBs and YMS/BS
stars behave differently in this colour range. The density of BHBs is
roughly constant, while the density of YMS/BS stars drops quickly
towards the lower $g-r$ values. In accordance with this pattern, the
YMS/BS contamination is highest in the left panel of
Figure~\ref{fig:grcuts}. There is a cloud of stars in the LMC disk at
$d_{\rm LMC}\sim10^{\circ}$ and $m-M\sim18.5$. However, no other
obvious overdensity is observable. In the middle and right panels,
where the number of contaminants is diminished, there appears to be a
clear BHB sequence stretching away from the LMC. This narrow BHB
overdensity shows a positive distance gradient with angular separation
from the dwarf. At around $d_{\rm LMC}\sim 50^{\circ}$, this dark band
reaches $m-M\sim19$, i.e. the heliocentric distance of the SMC
\citep[][]{SMCdist}. There are multiple other BHB overdensities beyond
$m-M\sim19$. However, given the faint magnitudes of the stars they are
composed of, it is difficult to ascertain the authenticity of all
structures present, especially those with $m-M > 20$. One striking
conclusion that follows from the picture presented above, is the
strong evidence for the extent of the stellar debris distribution as
far as $\sim50^{\circ}$ from the LMC.

\begin{figure}
  \centering
  \includegraphics[width=0.49\textwidth]{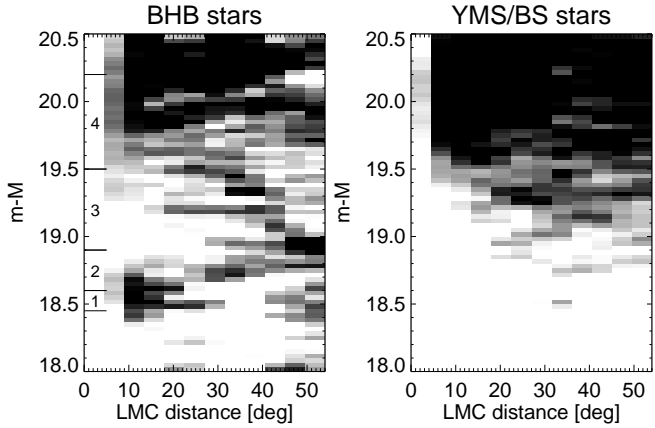}
  \caption[]{\small Line-of-sight clustering of BHBs as a function of
    the angular distance from the LMC. The maps are 12$\times$75
    pixels, and are smoothed with a Gaussian kernel with a FWHM of 1.7
    pixels. \textit{Left:} Density of the BHB stars in the plane of
    distance modulus and the distance from the LMC. As in
    Figure~\ref{fig:grcuts}, several dark sequences are clearly
    visible starting at $m-M\sim 18.5$, i.e. the LMC's heliocentric
    distance. Four ranges of $m-M$ explored in this work are shown on
    the left, and numbered from 1 to 4 for reference. \textit{Right:}
    Same as the middle panel but for the YMS/BS stars. In the distance
    modulus ranges selected, no obvious overdensity is discernible.}
   \label{fig:bhbdist}
\end{figure}

While $g-r<-0.075$ cut delivers a purer selection of BHBs, it also
reduces the number of tracers available. Given the fact that the
overdensity maps in the middle and the right panels of
Figure~\ref{fig:grcuts} are qualitatively the same, we choose to use
$g-r<-0.05$ limit as a default for the rest of the analysis. A more
detailed view of the sub-structures detected is given in
Figure~\ref{fig:bhbdist}. Here, left panel shows the same map of the
BHB overdensities as in Figure~\ref{fig:grcuts}, but zoomed-in onto
the $m-M$ range of relevance. Curiously, in this higher resolution
view, just next to the LMC, a narrow horizontal dark streak is
visible, running at a constant $m-M\sim18.5$ peeling off a broader BHB
sequence which slants up. This narrow sequence appears to come to an
abrupt stop around $d_{\rm LMC}\sim30^{\circ}$. As a comparison, the
right panel of the Figure presents the density of the YMS/BS stars in
the same coordinates as the previous panel. While no obvious coherent
patterns are visible at $m-M<20$, beyond that, the YMS/BS contamination
might grow substantially high. This is due to the combination of the
YMS/BS density hike and the rise of the $i-z$ error as a function of
apparent magnitude (as indicated in the top middle panel of
Figure~\ref{fig:bhb}).

\begin{figure*}
  \centering
  \includegraphics[width=0.98\textwidth]{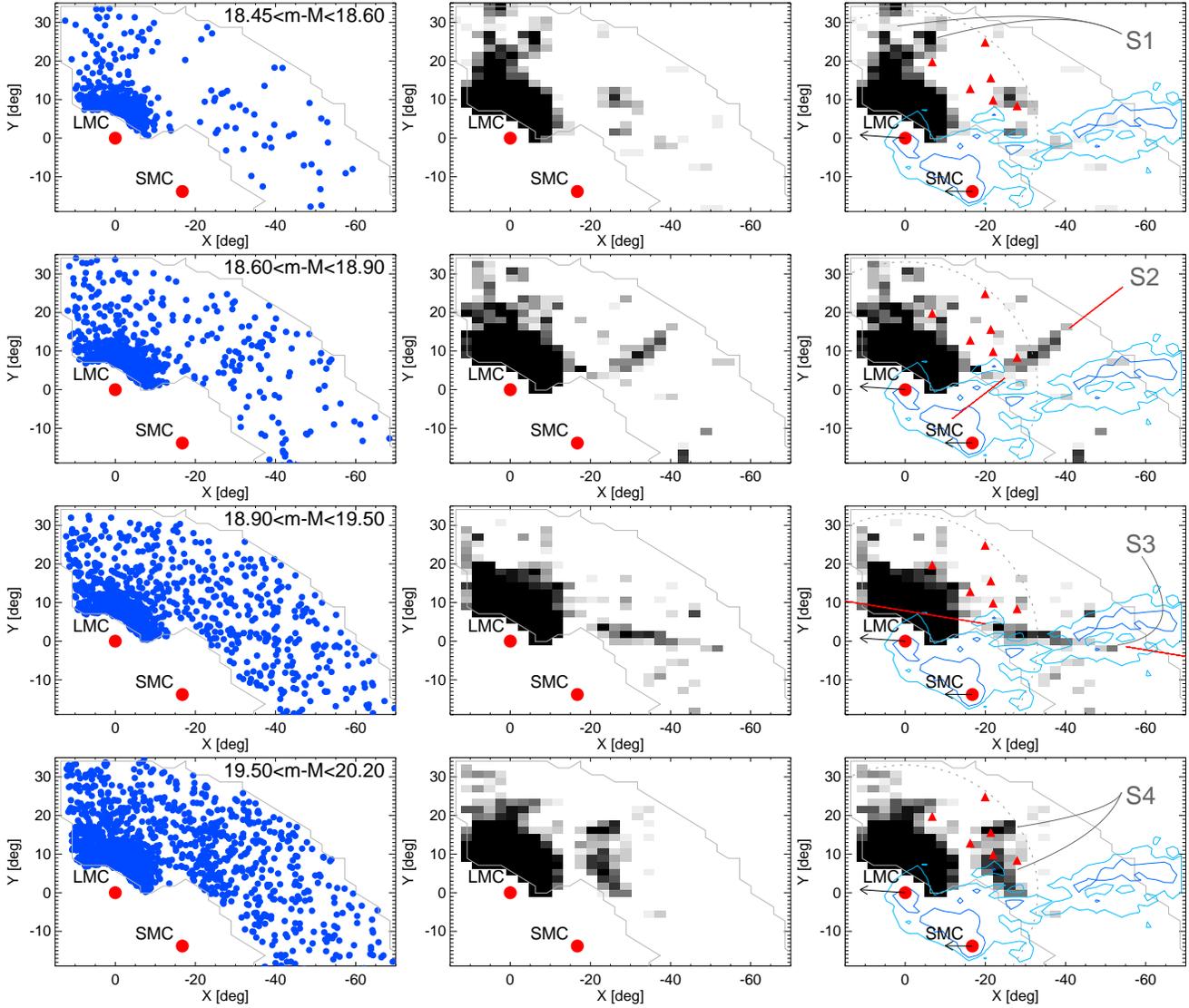}
  \caption[]{\small Distance slices of the LMC stellar
    halo. \textit{Left column:} Positions of the BHB stars in four
    distance modulus bins marked in the Figure~\ref{fig:bhbdist} and
    in the top right corner of the panel. \textit{Middle column:}
    Density of the BHB stars shown in the left column. The maps are
    30$\times$30 pixels, and are smoothed with a Gaussian kernel with
    a FWHM of 1.5 pixels. \textit{Right column:} As the middle column,
    but with auxiliary information over-plotted; this includes
    positions of the DES Year 1 satellites (red triangles) as well as
    the locations and the designations of the newly detected
    structures. \textit{1st row, from the top:} This is a narrow
    distance range centered around the LMC's $m-M=18.5$. A nearly
    vertical spray of BHB stars originating in the LMC is clearly
    visible, reaching all the way to the DES Year 1 footprint edge
    (marked with a light grey line), at $Y\sim35^{\circ}$. This
    sub-structure, dubbed S1, is also readily detectable in the
    density plots in the middle and the right panels. \textit{2nd
      row:} In this distance range, stopping short of the distance of
    the SMC, a messy plume of BHB stars is visible directly above the
    LMC, stretching out as far as $Y\sim20^{\circ}$, or perhaps even
    beyond. Additionally, between $X\sim -40^{\circ}$ and
    $X\sim-20^{\circ}$ a narrow stream-like structure is visible
    crossing the footprint edge to edge. This S2 stream is well fit by
    a great circle with the pole at $(\alpha, \delta)=(298\fdg5,
    177^{\circ})$, as indicated by the red lines shown in the right
    panel. \textit{3rd row:} As evidenced from the densely populated
    left panel, the background contamination with YMS/BS and WD stars,
    as well as the QSO has increased noticeably at this magnitudes. As
    well as the sub-structure in the northern region of the LMC's halo
    (as seen in the previous two distance bins), a cold nearly
    horizontal stream can be seen running from $X\sim-20^{\circ}$ to
    $X\sim -60^{\circ}$. The stream, dubbed S3, is approximate with
    the great circle with the pole at $(\alpha, \delta)=(250\fdg15,
    152\fdg35)$. \textit{4th row:} At these magnitudes, the background
    is the most severe, nonetheless, some interesting structures are
    still discernible. As confirmed by the density distribution, the
    two most prominent overdensities are i) a hook-like structures at
    $X\sim0^{\circ}$, $Y\sim20^{\circ}$ and ii) the S4 cloud of stars
    at $X\sim35^{\circ}$. on the sky, the S4 cloud appears to overlap
    several newly discovered DES satellites (red triangles). Arrows
    show the proper motion vectors of the LMC and the SMC, as measured
    by \citet{nitya2013} and corrected for the Solar reflex. Blue
    contours are the HI density of the Magellanic Stream as reported
    by \citet{nidever2008}. Positions on the sky with $D_{\rm
      LMC}=30^{\circ}$ are shown with grey dotted line.}
   \label{fig:zoom}
\end{figure*}

To reveal the locations of individual structures around the Magellanic
Clouds, the BHBs are split into four distance modulus bins (numbered
from 1 to 4, as indicated in the left panel of Fig~\ref{fig:bhbdist}),
namely: $18.45 < m-M<18.6$, $18.6 < m-M<18.9$, $18.9 < m-M<19.5$ and
$19.5 <m-M<20.2$. The spatial distributions (in gnomonic projection,
as described above) of stars in each bin are presented in
Figure~\ref{fig:zoom}, where only the portion of the footprint nearest
to the LMC is given, i.e. for stars with $Y>-20^{\circ}$. Each row
gives three views of the sub-structure: a scatter plot of all BHBs in
the corresponding distance bin, their density, and the density with
auxiliary information over-plotted. It is reassuring that all
structures described below are clearly visible in both scatter plots
and the density distributions. Starting from the top, the first row
shows the BHBs with the smallest distance modulus (corresponding to a
narrow range between 49 and 53 kpc), matching that of the LMC
itself. A striking, almost vertical tail of BHB stars at $X\sim
0^{\circ}$ is visible. The structure's geometry explains the abrupt
cut-off in the BHB sequence at $d_{\rm LMC}\sim30^{\circ}$ in the left
panel of Figure~\ref{fig:bhbdist}: this is where the stream, dubbed
S1, reaches the edge of the DES Year 1 footprint. Second row of the
Figure shows the density of slightly more distant BHBs, i.e. those
with $18.6 < m-M < 18.9$, corresponding to heliocentric distances
between 53 and 60 kpc. Again, around the LMC, a plume of BHBs is
visible, extending at least as far as $d_{\rm
  LMC}\sim20^{\circ}$. Moreover, a distinct narrow structure is
clearly noticeable further out, beyond $d_{\rm LMC}=30^{\circ}$. This
S2 stream runs diagonally from $(X,Y)=(-20^{\circ},0^{\circ})$ to
$(X,Y)=(-45^{\circ},15^{\circ})$, and appears, in fact, closer to the
SMC than the LMC. In the part of the sky covered by the DES Year 1
data, the stream's path follows the great circle with the pole at
$(\alpha, \delta)=(298\fdg5, 177^{\circ})$, as indicated by a red
line. The third (second from the bottom) row exhibits the yet narrower
and longer stream S3, traced by the BHBs with distances between 60 and
80 kpc. A great circle with the pole at $(\alpha, \delta)=(250\fdg15,
152\fdg35)$ seems to describe this stream well (see red line in the
right panel). Finally, the bottom row shows BHB candidates that are
possibly located as far as 110 kpc. Here, a curving tail of stars
emanating from the LMC and reaching as far as $Y\sim22^{\circ}$ is
discernible. Additionally, a cloud of stars (dubbed S4) with
$-30^{\circ} < X < -20^{\circ}$ and $0^{\circ} < Y < 20^{\circ}$ can
be clearly seen. Curiously, as indicated in the right panel of the
row, some of the recently discovered DES satellites appear coincident
with the cloud.

Also displayed in the right panel of Figure~\ref{fig:zoom} are arrows
representing the proper motion vectors of each of the Clouds. These
are the measurements of \citet{nitya2013}, corrected for the Solar
reflex motion using the Sun's velocity components with respect to the
local standard of rest (LSR) from \citet{ralf2010} and assuming
$V_{\rm LSR}=235$ km s$^{-1}$. Additionally, blue contours show the HI
density distribution in the Magellanic Stream as reported by
\citet{nidever2008}. Of the several stellar debris structures
described above, only the S3 stream runs parallel to the Clouds'
motion and overlaps sufficiently with the gaseous stream.

Figure~\ref{fig:radec} displays the distribution of the BHB density in
more familiar (yet more distorted) equatorial coordinates. Here, it is
possible to show the entire DES Year 1 footprint thanks to its simple
geometry in RA and Dec. The four vertically arranged panels once again
correspond to the 4 distance modulus bins under
consideration. Starting from the top, bins 1, 2 and 4 show clear signs
of stellar sub-structure directly above the LMC, the most prominent
being the S1 stream. Structures S2, S3 and S4 all show up in the
corresponding maps; however, additionally, there are hints of stellar
debris further away from the Magellanic Clouds, i.e. at
RA$=-10^{\circ}$ and RA$=-50^{\circ}$. Given that both of these
additional structures are most prominent in the density map
corresponding to the distance modulus 2, they are dubbed S2a and
S2b. Similar to Figure~\ref{fig:zoom}, approximate great circles are
shown for streams S2 and S3. Positions of DES Year 1 satellites and
the previously known Milky Way companions with distances less than 400
kpc are also marked with filled triangles and circles
correspondingly. As noted earlier, Hor 1 and 2, as well as Eri 3
appear to be coincident with the S4 debris cloud. Moreover, in the
equatorial view, the juxtaposition of Pho 2, Gru and Tuc 2 with the S2a
cloud appears plausible.

\subsection{Line-of-sight tomography of the debris}

\begin{figure*}
  \centering
  \includegraphics[width=0.98\textwidth]{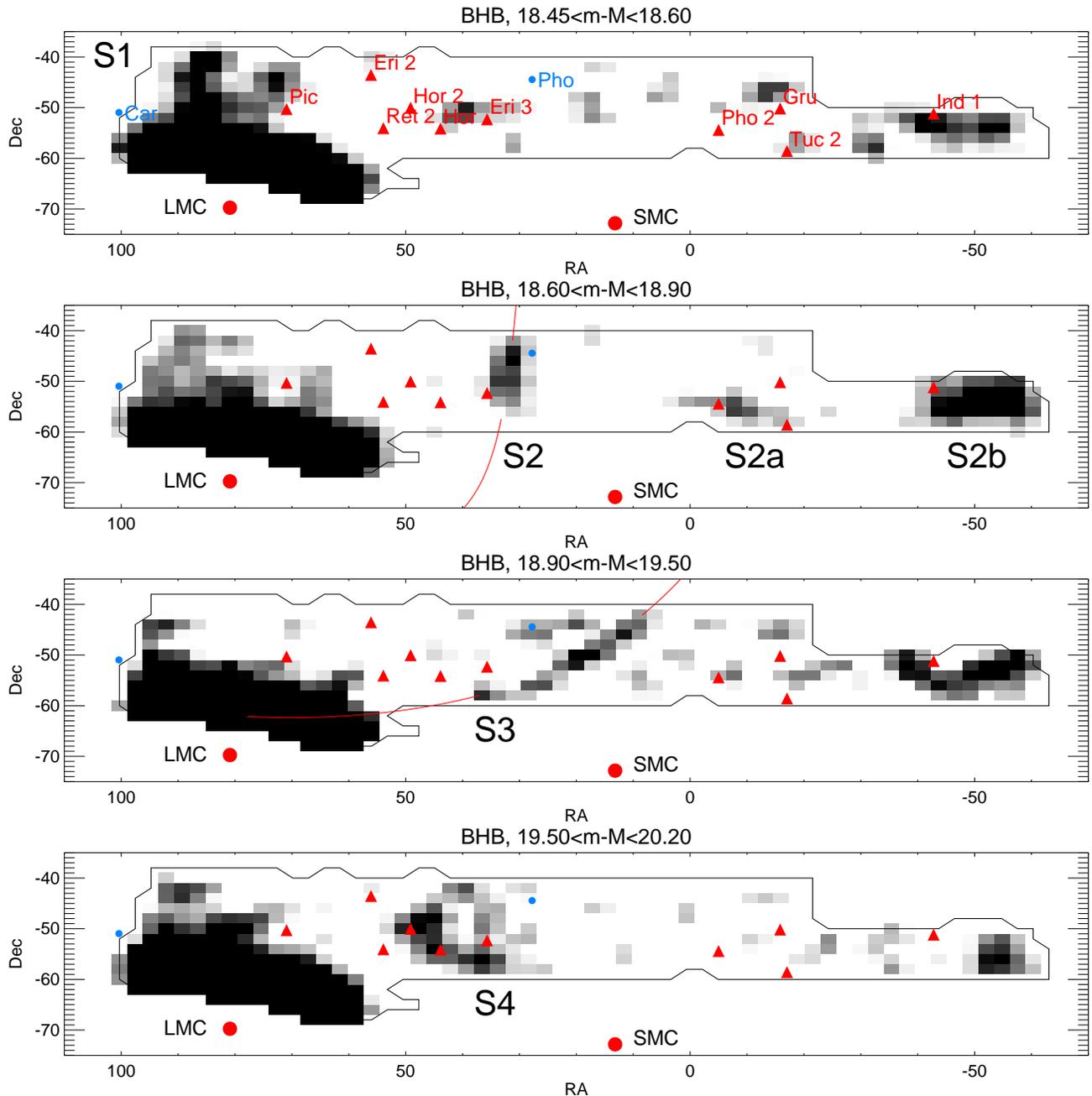}
  \caption[]{\small Magellanic stellar streams in equatorial
    coordinates. Top to bottom: density distributions of the BHB stars
    selected to lie in the four heliocentric distance bins described
    earlier.  The maps are 60$\times$25 pixels, and are smoothed with
    a Gaussian kernel with a FWHM of 2 pixels. Red filled circles show
    the locations of the LMC and the SMC, small blue filled circles
    mark the positions of the previously known satellites, while the
    red triangles correspond to the DES Year 1 satellites. Red lines
    show the great circle for streams S2 and S3. All structures
    identified in the gnomonic projection in the Figure~\ref{fig:zoom}
    are also present in these maps. Additionally, three more possible
    detections are noticeable, i.e. those at RA$\sim-10^{\circ}$ and
    RA$\sim -50^{\circ}$. Given that these structures are most
    prominent in the distribution of BHBs falling into the 2nd
    distance bins, they are dubbed S2a and S2b correspondingly. Note
    that Hor 1, Hor 2 and Eri 3 appear to overlap with the S2 cloud
    and its counterpart in the 1st distance modulus bin (top
    panel). Additionally, three other DES Year 1 satellites Pho 2, Gru
    and Tuc 2 seem to coincide with the S2a cloud.}
   \label{fig:radec}
\end{figure*}

\begin{figure*}
  \centering
  \includegraphics[width=0.98\textwidth]{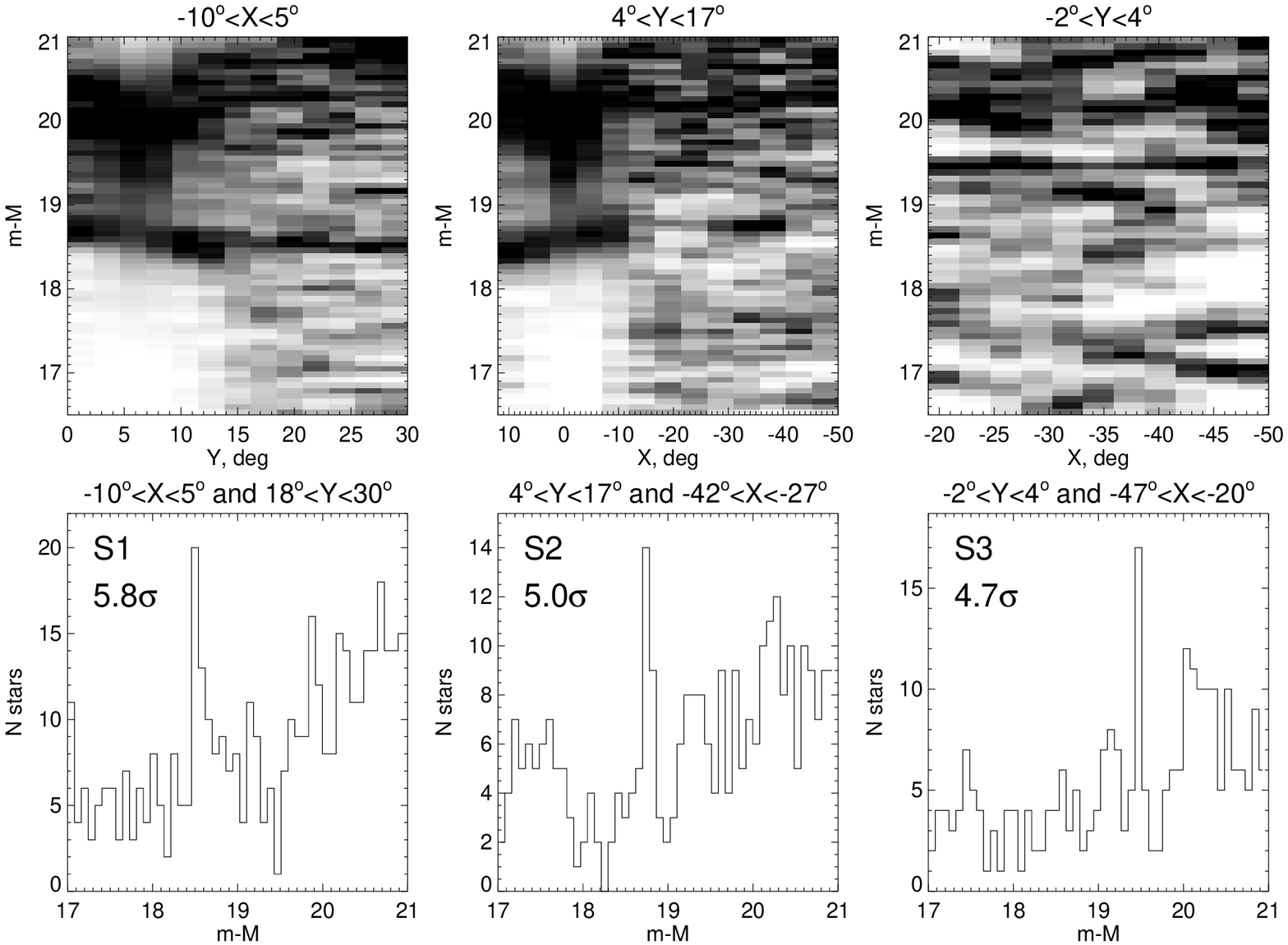}
  \caption[]{\small Line-of-sight tomography of structures S1, S2 and
    S3.  \textit{Top:} Density of BHBs in the space of distance
    modulus and one of the two spatial coordinates (of the gnomonic
    projection). The density is normalised per
    column. \textit{Bottom:} Histograms of distance modulus for each
    of the three structures under consideration. We also give the
    significance of each structure's peak in the top left corner in
    every bottom panel. \textit{Left:} This map is 13$\times$70
    pixels, and is smoothed with a Gaussian kernel with a FWHM of 1.6
    pixels. This is a vertical slice, crossing the LMC's northern
    regions. The LMC's disk dominates the range
    $0^{\circ}<Y<15^{\circ}$. A tilt of the disk is clearly visible
    reflected by a small but perceptible positive distance gradient as
    a function of Y. Note a narrow, horizontal overdensity connecting
    to the tilted disk feature around $Y\sim 14^{\circ}$. This is the
    S1 debris, located at $m-M\sim18.5$ and containing in excess of 20
    BHBs, as seen in the bottom panel. \textit{Middle:} This map is
    13$\times$70 pixels, and is smoothed with a Gaussian kernel with a
    FWHM of 1.5 pixels. This is a horizontal slice though the LMC. The
    familiar disk signature is visible at $m-M\sim18.5$, shadowed by a
    'V' shape cloud of the YMS/BS stars. As described in the text, we
    conjecture that the 'V' shape might be a result of the age
    gradient across the LMC's disk, with the central parts (at
    $X\sim0^{\circ}$) dominated by the younger (and hence brighter)
    stellar populations. The S2 stream is visible at
    $-40^{\circ}<X<-30^{\circ}$ and as shown in the bottom panel, its
    distance modulus distribution peaks around $m-m\sim18.75$ and
    contains $\sim$ 20 BHB stars. \textit{Right:} This map is
    11$\times$60 pixels, and is smoothed with a Gaussian kernel with a
    FWHM of 1.7 pixels.  Similar to the middle panel, i.e. a
    horizontal slice, but at lower $Y$. A very cold feature
    corresponding to the S3 stream is visible in both the top and the
    bottom panels. S3 is located at $m-M\sim 19.5$ and contains
    $\sim$15 BHB stars.}
   \label{fig:dmslice}
\end{figure*}

The most visible feature in all panels of both Figure~\ref{fig:zoom}
and Figure~\ref{fig:radec} is the strong overdensity of stars at the
LMC's location. However, as evident from Figure~\ref{fig:grcuts}, it
is likely that amongst these, only stars with $m-M<19$ are genuine
BHBs. The seemingly more distant objects are probably intrinsically
fainter stars (YMS/BS) in the LMC's disk. Luckily, as discussed at the
end of Section \ref{sec:data}, it is possible to distinguish between
the YMS/BS and BHB populations based on their clustering along the
line of sight. Thus Figure~\ref{fig:dmslice} sheds light on the
line-of-sight behaviour of the newly discovered stellar
sub-structures. In particular, panels in the top row of the Figure
give the greyscale BHB density maps in the plane of distance modulus
and one of the pair of the $X,Y$ coordinates (depending on the portion
of the sky under consideration). For example, top left panel shows the
distribution of the BHBs with $-10^{\circ} < X < 5^{\circ}$,
i.e. those located in the patch of the sky crossing the DES Year 1
footprint vertically around the position of the LMC. A thick and
mildly inclined slab of BHBs with $0^{\circ} < Y < 15^{\circ}$ and
centered around $m-M \sim 18.5$ is presumably the disk of the LMC. At
larger $m-M$, the disk BHBs are shadowed by the YMS/BS stars. The
density map is normalised vertically, i.e. each column is scaled by
the total number of stars it contains. This leads to a pronounced
underdensity at $m-M < 18.5$. There is also a noticeable gradient in
the distribution of the YMS/BS stars, vaguely reminiscent of the
letter V. Some of this pattern may be explained by the choice of the
density normalisation, but the principle cause of such an appearance
must be the gradient in the YMS/BS populations across the disk of the
LMC. The V-shape is more obvious in the next (top middle) panel of the
Figure, which presents the BHBs with $4^{\circ} < Y < 17^{\circ}$,
where the V's cusp can be seen pointing at $X=0^{\circ}$. The simplest
interpretation of such an apparent magnitude behaviour would be the
age gradient in the disk, with the younger and hence intrinsically
brighter MS stars located closer to the centre.

Also visible in the top left panel of Figure ~\ref{fig:dmslice}, a
much thinner strip of BHBs at a similar distance (i.e. $m-M\sim18.5$)
but extending beyond $Y\sim 14^{\circ}$ and reaching the edge of the
footprint at $Y\sim30^{\circ}$ - this is the S1 stream. The
line-of-sight distribution of the BHBs in S1 does not show any obvious
gradients in distance modulus as a function of $Y$. The bottom left
panel of the Figure displays the collapsed distribution of stars in
the stream, i.e. the BHBs with $-10^{\circ} < X < 5^{\circ}$ and
$18^{\circ} < Y < 30^{\circ}$. There is an obvious narrow peak in the
histogram with the maximum at $m-M=18.5$. Similarly, the two middle
panels present the results of the BHB tomography but for a patch of
the sky with $4^{\circ} < Y < 17^{\circ}$, i.e. a horizontal slice
through the LMC. As mentioned above, the LMC disk is apparent around
$m-M=18.5$ and $-10^{\circ} < Y < 10^{\circ}$. The S2 stream is
visible in the top middle panel at $27^{\circ} < X < 40^{\circ}$. It
does not connect to the LMC's disk feature and runs at a slightly
higher distance modulus, which is supported by the histogram in the
bottom middle panel, where the peak of the BHB count is
$m-M=18.75$. Finally, the top right panel presents the density of the
BHBs with $-2^{\circ} < Y < 4^{\circ}$. The S3 stream can be seen as a
narrow dark streak running across the entire figure from
$X=-21^{\circ}$ to $X=-50^{\circ}$. According to the corresponding
bottom panel, the peak distance modulus of S3 is at $m-M=19.5$. For
every structure discussed above, the significance of the BHB
overdensity along the line of sight is given in the top left corner of
the bottom row panels of Figure~\ref{fig:dmslice}. We have also
examined similar distance modulus slices for the structures S4, S2a
and S2b. While there are peaks at around $m-M\sim18.8$ for S2a and
S2b, these are not as statistically significant as those for S1, S2
and S3 discussed above. Additionally, no obvious narrow peak is
detected for S4; in fact, S4 could plausibly be produced by the
overdensity of YMS/BS stars at distances similar to that of the LMC.

\subsection{Notes on individual structures}

In the absence of kinematical and chemical information, it is
difficult to pinpoint the origin of the structures
detected. Nonetheless, below, we hazard a guess based on the shape and
the orientation of the debris distributions on the sky. We found our
speculation on the fact that the width of the stellar stream reflects
primarily the mass of its progenitor, and even though it is also a
function of the stream's dynamical age as well as the asphericity of
the host gravitational potential, these effects are secondary
\citep[see e.g.][]{johnston1998,erkal2015}. For example, in the Milky
Way, globular cluster streams are only $\sim$100 pc accross \citep[see
  e.g.][]{pal5,gd1,atlas}, while those originating in dwarf galaxies
have more substantial widths of several kpc \citep[see
  e.g.][]{orphan,Ko12}. On the other hand, stellar streams produced as
a result of the interaction between the LMC and the SMC show a variety
of cross-sections, from $\sim$2 to 10-15 kpc
\citep[see][]{besla2013,diaz2012}, but never appear to be much
narrower than 1 kpc.

\subsubsection{S1 debris, LMC's stellar halo and the S4a candidate stream}

\begin{figure*}
  \centering
  \includegraphics[width=0.98\textwidth]{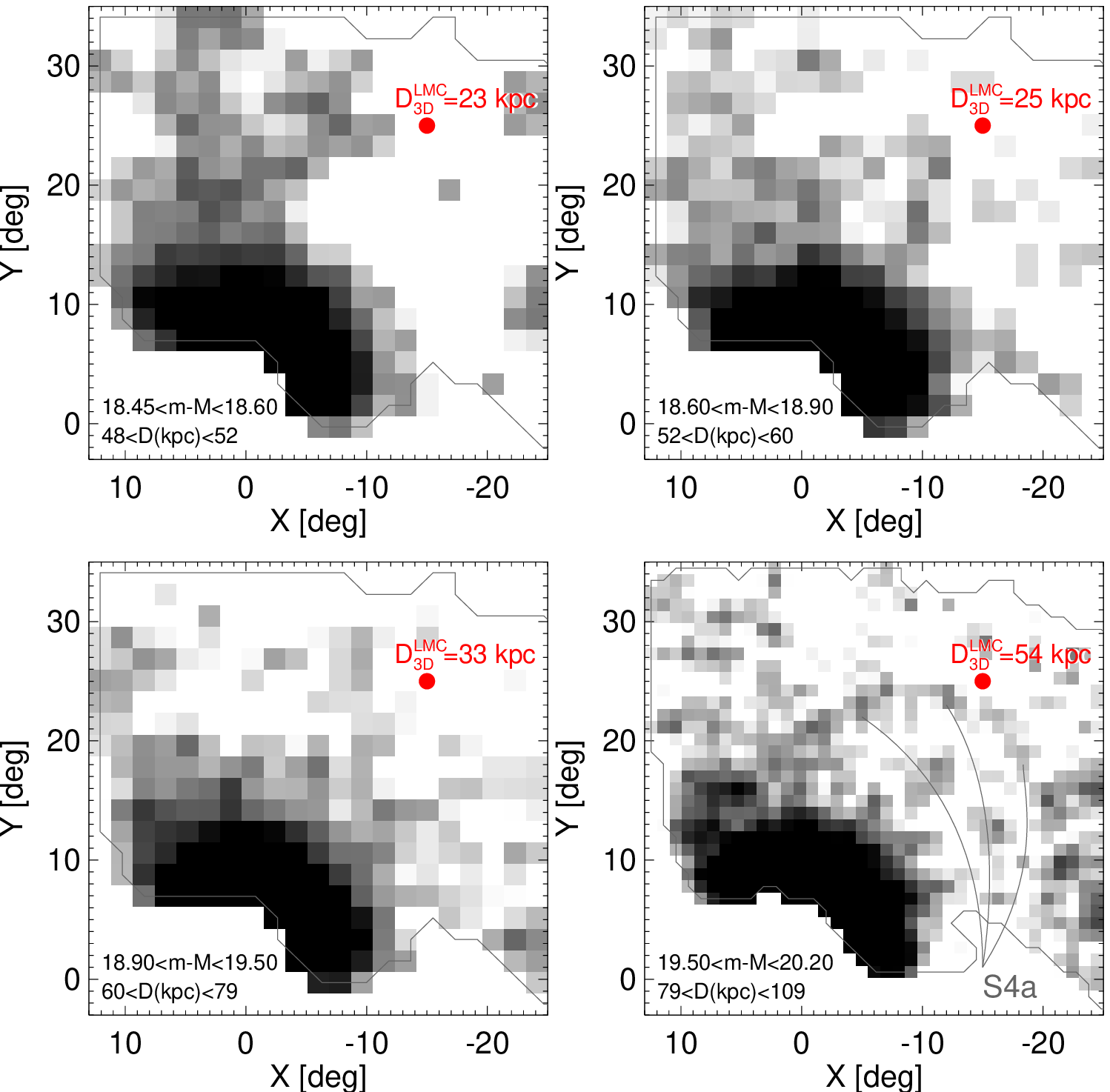}
  \caption[]{\small Zoomed-in view of the Northern portion of the LMC
    stellar halo. Each panel presents the density of BHB selected to
    have distances in the four distance modulus ranges (given in the
    bottom left corner). In each case, debris reaching at least
    $Y=20^{\circ}$ are visible. The last panel (bottom right),
    corresponding the largest $m-M$, also appears to contain a
    tentative stream-like structure dubbed S4a. Note however, that at
    this faint magnitudes the background/foreground contamination is
    most severe. Red filled circle marks the location of a reference
    point with $X=-15^{\circ}$ and $Y=25^{\circ}$. In each panel, the
    number next to the red circle gives the 3D distance in kpc of the
    reference point from the LMC's centre, assuming the mean
    line-of-sight distance to the debris for each $m-M$ range. The
    maximal separation between the LMC's centre and the stellar halo
    sub-structure varies from 25 to 50 kpc.}
   \label{fig:lmczoom}
\end{figure*}

Figure~\ref{fig:lmczoom} zooms in onto the stellar debris in the
immediate vicinity of the LMC. The four panels of the Figure
correspond to the four distance modulus bins introduced
above. Remarkably, there is plenty of evidence of stellar halo
sub-structure in all four panels. Displaying the density distribution
of the BHBs at around the LMC's distance, the top left panel emphasizes
the dramatic appearance of the S1 structure. Moving further along the
line of sight, the top right panel reveals what might appear like tidal
debris loops rising above $Y\sim20^{\circ}$. The bottom left panel appears
the smoothest and most symmetric of the four. Finally, the bottom right
panel, corresponding to BHB distances between 80 and 110 kpc, uncovers
an arch of a narrow stream-like structure. Dubbed S4a, following the
same nomenclature, its base is at $X=5^{\circ}, Y=15^{\circ}$ and it
is easily traced as far as $X=-18^{\circ}, Y=18^{\circ}$. In fact, the
denser and the closest to the LMC part of the stream (around
$X\sim0^{\circ}$) can also be seen in Figures~\ref{fig:zoom} and
\ref{fig:radec}. Note though, that given the low number of stars
contributing to the feature and the growing levels of contamination at
these apparent magnitudes, the S4a structure can be a mere chance
alignment of BHB candidate stars. If confirmed as a genuine stream,
however, S4a can deliver powerful constraints on the properties of the
dark matter distribution in the LMC.

The extent of the LMC's stellar halo can be gauged by comparing the
distribution of the BHB sub-structure as revealed in each panel of
Figure~\ref{fig:lmczoom} with the reference point at $X=-15^{\circ}$,
$Y=25^{\circ}$ (red filled circle). The 3D separation between the
LMC's centre and the reference point is calculated assuming the mean
heliocentric distance in each $m-M$ bin. Depending on the distance
range considered, the stellar halo sub-structure traced by the BHBs in
the DES Year 1 data reaches distances between 25 and 50 kpc away from
the LMC's centre.

\subsubsection{S2 stream}

\begin{figure}
  \centering
  \includegraphics[width=0.48\textwidth]{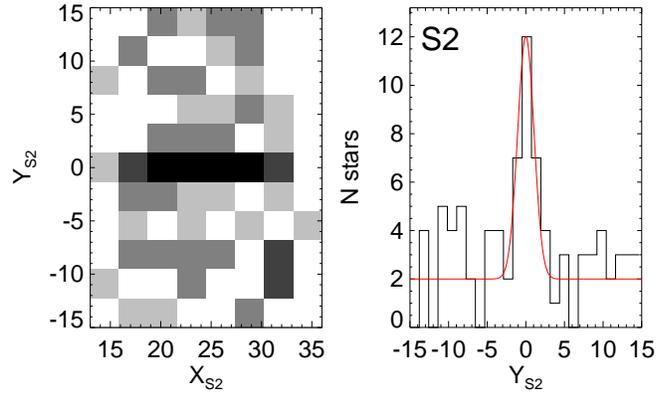}
  \caption[]{\small S2 stream. BHBs with distances $18.6 <m-M < 18.9$
    are shown in the coordinate system aligned with the stream
    $X_{S2},Y_{S2}$, i.e. with the great circle with the pole at
    $(\alpha, \delta)=(298\fdg5, 177^{\circ})$. The origin of the
    $X_{S2}$ axis is the projection of the SMC's position onto the
    great circle. {\it Left:} Density of stars in
    $X_{S2},Y_{S2}$. {\it Right:} Across-stream density profile of the
    stream as traced by the stars with distances as described above
    and $15^{\circ} < X_{S2} < 35^{\circ}$. To guide the eye, red line
    shows a Gaussian profile with $\sigma=1\fdg5$}
   \label{fig:s2profile}
\end{figure}

The S2 stream crosses the DES Year 1 footprint almost vertically along
the Dec direction (see Figure~\ref{fig:radec}). The stream's track is
well represented with the great circle with the pole at $(\alpha,
\delta)=(298\fdg5, 177^{\circ})$. As clear from Figure~\ref{fig:zoom},
the stream passes between the two Magellanic Clouds, almost
perpendicular to the line connecting the two galaxies, somewhat closer
to the SMC. The stream's heliocentric distance is approximately in
between the LMC's and the SMC's, at 56 kpc. As can be gleaned from the
middle top panel of Figure~\ref{fig:dmslice}§, there is a hint of a
positive distance gradient in the trailing direction of the LMC's
motion, i.e. decreasing $X$. Figure~\ref{fig:s2profile} gives the view
of S2 in the coordinate system aligned with the stream. As
demonstrated in the left panel of the Figure, the current DES
footprint covers the S2 debris in the range $15^{\circ} < X_{S2} <
33^{\circ}$. At the distance of S2, this translates into 15 and 33 kpc
away from the line connecting the LMC and the SMC. The right panel of
Figure~\ref{fig:s2profile} shows the across-stream profile of the S2
debris. As indicated by the overlaid red line, the stream's density
can be approximated with a Gaussian profile with FWHM$\sim3\fdg5$, or
at the S2's distance, $\sim3.5$ kpc. Given the fact that the width of
the stream is substantial, but seemingly smaller than what is expected
for the LMC or the SMC, it is perhaps possible the stream originates
from a dwarf galaxy disrupting in the tidal field of the Magellanic
system.  In fact, the aspect ratio of the detected piece of debris is
only 5-to-1, therefore rather than a section of a narrow and long
stream, S2 could be the remnant of a tidally stretched dwarf
galaxy. Note, however, that, if S2 is indeed a stream that orbits one
of or both the Magellanic Clouds, it should not, in principle, be
aligned with a great circle everywhere on the celestial
sphere. Therefore, it is also possible to interpret the S2 debris as a
disruption of a system that used to be part of the Magellanic Family
in the past and is now orbiting the Milky Way.

\subsubsection{S3 stream}

\begin{figure}
  \centering
  \includegraphics[width=0.48\textwidth]{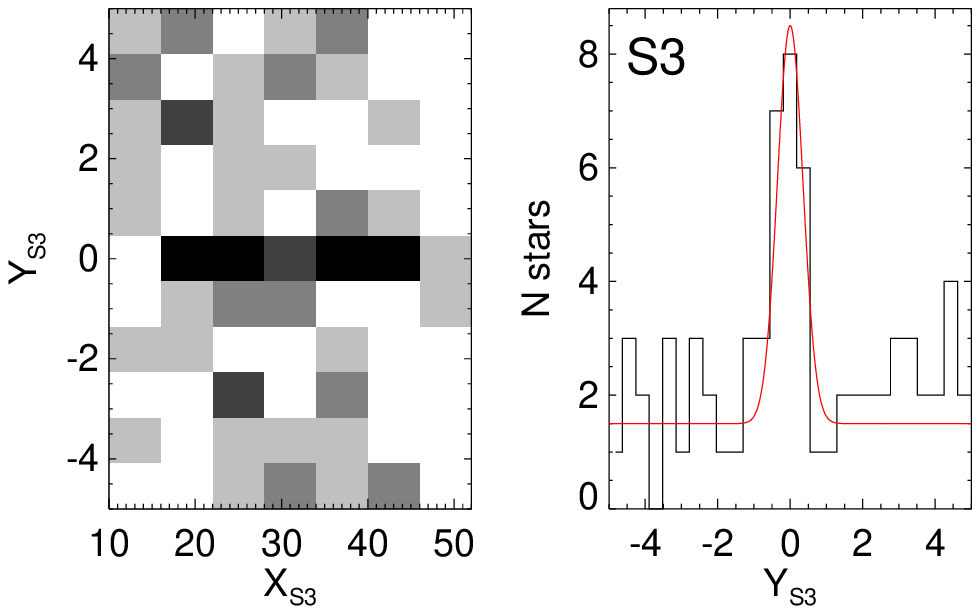}
  \caption[]{\small S3 stream. BHBs with distances $19.3 <m-M < 19.6$
    are shown in the coordinate system aligned with the stream
    $X_{S3},Y_{S3}$, i.e. with the great circle with the pole at
    $(\alpha, \delta)=(250\fdg15, 152\fdg35)$. The origin of
    the $X_{S3}$ axis is the projection of the LMC's position onto the
    great circle. {\it Left:} Density of stars in
    $X_{S3},Y_{S3}$. {\it Right:} Across-stream density profile of the
    stream as traced by the stars with distances as described above
    and $15^{\circ} < X_{S3} < 51^{\circ}$. To guide the eye, red line
    shows a Gaussian profile with $\sigma=0\fdg5$}
   \label{fig:s3profile}
\end{figure}

The S3 stream, running along the great circle with the pole at
$(\alpha, \delta)=(250\fdg15, 152\fdg35)$, points almost exactly
towards the LMC (see Figures~\ref{fig:zoom} and \ref{fig:radec}). As
evidenced by the right column of Figure~\ref{fig:dmslice}, it is
located at the heliocentric distance of $\sim 80$ kpc, i.e. further
away than both of the Magellanic Clouds. Figure~\ref{fig:s3profile}
presents the view of S3 in the coordinate system aligned with the
stream, $X_{S3}, Y_{S3}$. The apparent width of S3 is clearly quite
much smaller as compared to that of S2, as discussed above. The right
panel of the Figure shows the across-stream profile, which appears to
be only $1\fdg2$ across, or 1.6 kpc at the distance of the stream. It
is not impossible that the stream is actually even narrower than what
is indicated by the handful of BHBs detected in the DES data: for
example, if the stream's track is not well-described by a great
circle, our estimate of the width could be biased high. Given the
uncertainty in the stream's width, it is difficult to discern with
confidence whether its progenitor was a dwarf galaxy or a globular
cluster. As judged by the orientation of its orbit, it might appear to
be in the process of being accreted onto the LMC. However, in 3D, the
point of its closest approach to the LMC is $\sim 20$ kpc. If this is
indeed the stream's pericentre, it is unlikely that the progenitor
could have suffered any significant tidal stripping this far
out. Moreover, the caution in the interpretation of the great circle
alignment as raised above applies to the S3 stream as well. Overall,
the least unlikely scenario seems to be the one in which the stream's
progenitor fell into the MW as part of a loose group associated with
the LMC. Finally, given the S3's alignment with the Clouds' proper
motion and the overlap with the gaseous MS, it can plausibly be a
product of the LMC-SMC interaction.


\section{Discussion and Conclusions}
\label{sec:disc}

Most of the previous attempts to find tidally stripped stars at large
distances around the Magellanic Clouds have been disappointing as no
clear presence of the stellar counterpart to the prominent gaseous
stream has been reported. Additionally, only fragmentary glimpses of
the alleged LMC's stellar halo have been seen; in fact, many of these
detections could be linked directly to the galaxy's disk, rather than
the genuine halo. In a bid to produce the final verdict as to the
existence of a substantial amount of stellar debris in the halo of the
Clouds, we use the deep, wide and continuous coverage of the
Magellanic environs provided by the DES Year 1 imaging data. Our
tracer of choice is Blue Horizontal Branch stars; to our knowledge,
these have not yet been used to study the Clouds. In this paper, we
demonstrate that with the help of the $griz$ photometry, highly
complete samples of BHBs can be selected, suffering only moderate
levels of contamination from QSOs, WDs and YMS/BS stars. The crucial
part in the successful BHB identification is played by the $i-z$
colour, which gauges the flux excess due to differences in the shapes
of the Paschen lines in the near-IR part of the spectra of stars with
low and high surface gravity. As shown in the top middle panel of
Figure~\ref{fig:bhb}, the photometric $i-z$ error only starts to blow
up at $g\sim21$, which gives us confidence that thus selected BHBs
can trace structures robustly as far as $m-M\sim20$, i.e. at least 100
kpc.

Armed with a large sample of BHBs detected across several thousands of
square degrees, we have exposed -- immense in scale and rich in
sub-structure -- the stellar halo of the Magellanic system. As
Figures~\ref{fig:grcuts} and ~\ref{fig:bhbdist} demonstrate, around
the Clouds, multiple BHB agglomerations exist, stretching from edge to
edge of the DES Year 1 footprint, ($\sim50^{\circ}$ from the
LMC). These sub-structures are clearly dispersed over an enormous
volume of tens of kpc: not only they show a complex behaviour as a
function of the position on the sky, they also are intertwined along
the line of sight, starting from the heliocentric distance of the LMC
at $\sim50$ kpc and continuing beyond 100 kpc. Note that it is not
surprising that we predominantly see stars with distances larger than
that of the LMC. This could simply be due to the fact that the DES
Year 1 footprint covers the area of the sky predicted to be populated
with the trailing (and hence more distant) debris.

If the BHBs are grouped according to their heliocentric distance,
overdense regions corresponding to individual stellar streams are
immediately obvious. We detect at least four distinct Magellanic
stellar halo structures labeled from S1 to S4 according to the
distance modulus bin they occupy. An overview of their properties can
be found in Figure~\ref{fig:zoom}. These discoveries represent the
very first census of the Magellanic stellar halo
sub-structure. Different in shape, extent and luminosity, each of
these clearly, deserves its own detailed analysis, which due to the
obvious lack of space, can not be part of this Paper. Nevertheless, a
few important highlights are worth mentioning. For example, S1 is the
nearest of the structures discovered and is the only one unambiguously
connected to the LMC. Even though it stretches to at least
$\sim30^{\circ}$ from the LMC, its true extent is not known since at
its most distant point, the stream is cut-off by the DES Year 1
footprint. As far as we can see, S1 does not appear to be directly
related to the 10 kpc stellar stream recently reported by
\citet{Mackey2015}. Moreover, as evidenced by the top left panel of
Figure~\ref{fig:dmslice}, around the location where S1 decouples from
the LMC disk BHB sequence, its debris do not seem to follow the
apparent disk's distance gradient (note the discontinuity at
$Y\sim15^{\circ}$). This might imply a different origin for S1,
perhaps in the LMC's stellar halo.


We have also discovered two narrow stellar streams at distances beyond
$30^{\circ}$ from the LMC. S2 and S3 are also further away along the
line of sight, at 56 kpc and 80 kpc correspondingly. The FWHM of S2 is
$\sim 3\fdg 5$ and thus it is clearly the product of a disruption of a
dwarf galaxy. S3 is even narrower, with FWHM$\sim 1\fdg2$. However,
the stream's width might only be an upper bound as the true trajectory
of S3 is unclear. In fact, both S2 and S3, at least their parts
visible in the DES Year 1 data, run approximately along great
circles. This could be the result of the incomplete view of the
streams, to be rectified when more area near the Clouds is
available. It is, however, not impossible that S2 and S3 indeed do not
curve around either of the galaxies and are now orbiting the Milky Way
instead. Interestingly, both S2 and S3 appear to be approximately
aligned with the proper motion of the Clouds. Moreover, S3 overlaps
with the Magellanic Stream on the sky. Could these therefore be the
stellar counterparts of the broader gaseous MS? It is difficult to
rule the hypothesis out, given that narrow streams (like S2 and S3)
can actually be produced in the disruption of a massive dwarf galaxy,
provided the progenitor is not a pressure supported system but a cold
rotating disk \citep[see e.g.][]{donghia2009,jorge2010}. In fact,
  both the LMC and the SMC seem to fit the bill.

Amorphous in appearance, the S4 and S2a clouds are more difficult to
interpret without the help of a detailed numerical simulation of the
Clouds' disruption. It is curious, nonetheless, that several of the
recently discovered DES satellites seem to coincide with either of
these in both position on the sky and the distance along the line of
sight. Figure~\ref{fig:lmczoom} zooms in onto the regions of the sky
between 10$^{\circ}$ and 30$^{\circ}$ from the LMC. This is the very
first sweeping panorama of the Cloud's stellar halo. Even in this
limited view, it appears gigantic in size: overdensities are detected
as far as $\sim30^{\circ}$ from the LMC, corresponding to 3D distances
from 25 kpc to 50 kpc. We therefore hypothesize that the break in the
LMC's stellar density at 13.5 kpc from its centre reported by
\citet{balbinot2015a} is not a sign of the tidal truncation but rather
of the transition from the (perturbed) disk to the stellar halo. Note
that none of the structures presented here overlap in 3D with the
Phoenix stream and the Eridanus-Phoenix cloud recently discovered in
the same region of the sky, but at much lower distances by
\citet{balbinot2015b} and \citet{li2015}.

Faced with the wealth and the expanse of the stellar halo
sub-structure discovered, it is prudent to ask whether the LMC could
in fact be much more massive than has been assumed until
recently. Currently, the mass distribution in the LMC has only been
probed out to $\sim$9 kpc from the dwarf's centre, where according to
\citet{vdm2002} it adds up to $\sim9\times10^9 M_{\odot}$. Therefore,
many of the numerical MS models assumed a total of $\sim 10^{10}
M_{\odot}$ for the LMC's mass \citep[see
  e.g.][]{connors2006,diaz2012}. However, several new - albeit
indirect - clues have been identified that imply that the total mass
of the Cloud could be at least an order of magnitude higher. For
example, \citet{Besla2012} found that the infall of an LMC with a
total mass of $\sim 2 \times 10^{11} M_{\odot}$ could explain many of
the key properties of the Magellanic Stream. On the other hand, a
simulation based on low mass LMC appears equally successful \citep[see
  e.g.][]{diaz2012}. Additionally, \citet{deason2015} confirmed with a
suite of high-resolution Cosmological N-body simulations, that a
recent accretion of a massive, i.e. $>10^{11} M_{\odot}$ LMC would
account for a large number of satellites detected in the DES
footprint. Note, however, that according to the earlier study of
\citet{Sales2011} even a much smaller mass LMC, i.e. that with
$3\times 10^{10} M_{\odot}$, could drag an impressive amount of
low-mass substructure into the MW halo. Importantly, in light of their
relatively high velocities, keeping the Clouds a bound pair for a
significant fraction of the Hubble time also requires an LMC with a
mass exceeding $10^{11} M_{\odot}$ \citep[see][]{nitya2013}. Finally,
using an extended timing argument calculation, although not taking the
presence of the M33 into account, \citet{jorge2015} estimated that the
LMC's mass could easily be $2.5 \times 10^{11}$ M$_{\odot}$.

Notwithstanding the large number of numerical experiments dedicated to
the analysis of the Magellanic infall, the only channel of its stellar
halo formation that has been explored so far is that due to the
LMC-SMC interaction. Most models produce plenty of the SMC stellar
spraying around the LMC. However, even in the recent treatise of
\citet{Besla2012} and \citet{diaz2012}, very little SMC stellar debris
can be seen in the area of the sky discussed here, i.e. $Y>0^{\circ}$,
and almost none at $Y>15^{\circ}$. Could any of the stellar
sub-structure identified here be due to accretion and tidal disruption
of smaller mass satellites? According to the most recent abundance
matching projections, some $\sim 50$ dwarf satellites could have
orbited the LMC \citep[see e.g.][]{deason2015}. The number of star
clusters with apo-centres beyond 10 kpc from the Cloud is more
difficult to estimate. The question of such high-orbit cluster
production is linked to the poorly understood specific frequency of
GCs in the dwarf satellites of the LMC. The possibility of cluster
formation in the intra-cloud space has not been ruled out either, nor
has the exchange of star clusters between the LMC and the SMC. If
numerous star clusters and dwarf galaxies did orbit the LMC, it is
perhaps possible that a fraction of these could be tidally stripped to
produce at least some of the stellar halo observed in the DES
data. This, of course, would require favourable orbits, i.e. such that
would bring the satellite into the densest regions of the LMC, well
within the inner 10 kpc.

What is the orbital history of the Clouds? How did their interaction
proceed? What was the satellite luminosity function in the LMC? What
is the mass profile of the LMC? These questions can now be addressed
through a combination of the deep imaging and spectroscopic follow up
of the tidal debris uncovered in this work.

\section*{Acknowledgments}
The authors wish to thank Wyn Evans, Mike Irwin, Denis Erkal, Jonathan
Diaz and Natalia Mora-Sitj\`a for illuminating discussions that helped
to improve this manuscript. We are grateful to the anonymous referee
for the helpful comments they have provided. The research leading to
these results has received funding from the European Research Council
under the European Union's Seventh Framework Programme (FP/2007-2013)
/ ERC Grant Agreement n. 308024. V.B. and S.K. acknowledge financial
support from the ERC. This research was made possible through the use
of the AAVSO Photometric All-Sky Survey (APASS), funded by the Robert
Martin Ayers Sciences Fund.

\label{lastpage}

\end{document}